\documentclass[aps,prl,twocolumn,nofootinbib,groupedaddress]{revtex4-1}
\usepackage{amsmath,amsfonts,amssymb,bbm,epsfig}

\def\be{\begin{equation}}
\def\ee{\end{equation}}
\def\bq{\begin{equation}}
\def\eq{\end{equation}}
\def\bqa{\begin{eqnarray}}
\def\eqa{\end{eqnarray}}
\def\roughly#1{\mathrel{\raise.3ex
\hbox{$#1$\kern-.75em\lower1ex\hbox{$\sim$}}}}
\def\lsim{\roughly<}
\def\gsim{\roughly>}

\newcommand{\rb}[1]{\raisebox{1.5ex}[-1.5ex]{#1}}

\begin{document}


\title{Color-Dipole Picture versus Hard Pomeron\\
in Deep Inelastic Scattering}


\author{Dieter Schildknecht}
\affiliation{Fakult\"at f\"ur Physik, Universit\"at Bielefeld\\
D-33501 Bielefeld, Germany\\
and\\
Max-Planck-Institut f\"ur Physik (Werner-Heisenberg-Institut)\\
F\"ohringer Ring 6, D-80805 M\"unchen, Germany}


\date{\today}

\begin{abstract}
For photon virtualities of $Q^2 \gsim 20 {\rm GeV}^2$, the results of the (hard
Pomeron) fit in the tensor-Pomeron model in terms of the variables $Q^2$ and the virtual-photon-proton energy squared, $W^2$, provide empirical evidence for
the validity of the Color Dipole Picture (CDP).  Consistency
of the CDP with the perturbative QCD (pQCD) improved parton model implies
the prediction of $C_2 = \epsilon_0 \cong 0.30$ for the exponent $C_2$ of the
energy-squared dependence in agreement with the results of the fits.
For $Q^2 \lsim 20 GeV^2$,
the CDP yields a parameter-free smooth transition from $Q^2 \gsim 20
{\rm GeV}^2$
to $Q^2 = 0$ photoproduction,
in distinction from the tensor-Pomeron model that relies on the additional
parameter quantifying the intercept of the soft-Pomeron trajectory.
\end{abstract}

\pacs{}

\maketitle


\section{Introduction}
The recent article \cite{Ewerz} by Britzger, Ewerz, Glazow, Nachtmann and Schmidt revives the two-Pomeron-plus-Reggeon approach from ref. \cite{Donnachie} by introducing tensor couplings of the hard Pomeron, $\mathbbm{P}_0$,
the soft Pomeron, $\mathbbm{P}_1$, and the $f_{2R}$ Reggeon at the photon and proton vertices, $\gamma^* \gamma^*(\mathbbm{P}_0, \mathbbm{P}_1, f_{2R})$ and $pp (\mathbbm{P}_0, \mathbbm{P}_1, f_{2R})$, to fit the experimental data on deep-inelastic lepton-nucleon scattering (DIS) at low values of the Bjorken variable
$x \simeq Q^2/W^2 < 0.01$, including the $Q^2 = 0$ photoproduction limit\footnote{In standard notation, $Q^2$ and $W^2$ denote the photon virtuality and the square of the photon-proton center-of-mass energy.}.

Regge theory, by generalization from hadronic interactions, assumes a power-law $W$ dependence for the photoabsorption cross section, as $(W^2)^{\alpha_j (0)-1}$, where the three \cite{Donnachie} Regge intercepts, $\alpha_0 (0) = 1 + \epsilon_0, \alpha_1 (0) = 1 + \epsilon_1$ and $\alpha_2 (0)$, are free parameters to be extracted from a fit to the experimental data. The form of the $Q^2$ dependence of the $\gamma^* \gamma^* (\mathbbm{P}_0, \mathbbm{P}_1, f_{2R})$ vertex contributions not being fixed by Regge theory, in ref. \cite{Ewerz} the $Q^2$ dependence is model-independently parametrized by smooth functions in terms of splines.
Altogether the fit \cite{Ewerz} contains 25 free parameters, namely the 3 Regge intercepts, 5 parameters for the
$\gamma^* \gamma^* (\mathbbm{P}_0, \mathbbm{P}_1, f_{2R})$ couplings at $Q^2 = 0$, and 17 parameters for a model-independent description of the $Q^2$ dependence in terms of spline functions.

In the present article, we confront the interpretation of the DIS experimental data in Regge theory with their representation in the Color-Dipole Picture (CDP). We proceed in three steps. In a first step, we concentrate on the region of large $Q^2 \gsim 20 {\rm GeV}^2$, in Regge theory associated with the hard Pomeron $\mathbbm{P}$ with intercept $\alpha_0 (0)$. In the second step, we show that consistency of the CDP and the perturbative-QCD-improved parton model implies a successful prediction for the numerical value of the exponent of the
$W^2$ dependence, $\epsilon_0 = C_2 \cong 0.30$. In the third step, we consider the transition to low $Q^2$, including the $Q^2 = 0$ limit of photoproduction, in Regge theory associated with the soft Pomeron with intercept $\alpha_1 (0)$.

It will turn out that the model-independent 17-parameter fit of the $Q^2$ dependence in the tensor-Pomeron approach for $Q^2 \gsim 20 {\rm GeV}^2$ leads to, and confirms,
the {\it parameter-free prediction} of the
$Q^2$-dependence of the CDP, wherein the photoabsorption cross section, $\sigma_{\gamma^*p}
(W^2,Q^2)$, depends on the low-x scaling variable $\eta^{-1} = \Lambda^2_{sat} (W^2) (Q^2+m^2_0)^{-1}$,
i.e. $\sigma_{\gamma^*p} (W^2,Q^2) = \sigma_{\gamma^*p} (\eta (W^2,Q^2))$,
with $\Lambda^2_{sat} (W^2) \propto (W^2)^{C_2}$ and $C_2 = \epsilon_0$. Specifically, for 
$Q^2 \gsim 20 {\rm GeV}^2$, in the CDP, $\sigma_{\gamma^*p} \propto \eta^{-1} = \Lambda^2_{sat} (W^2)/Q^2$,
as confirmed by the tensor-Pomeron-fit result.

\section{Large Values of $\mathbf{Q^2 \gsim 20 {\rm \bf GeV}^2}$}.
For large values of $Q^2$, the total photoabsorption cross section in the Regge approach is determined \cite{Ewerz} by the hard-Pomeron contribution.
According to formula (6.1) in ref. \cite{Ewerz}, the fit led to the result
\be
\sigma_{\gamma^*p} (W^2,Q^2) \propto \hat{b}_0 (Q^2) (W^2)^{\epsilon_0},
\label{1}
\ee
where
\be
\epsilon_0 = 0.3008 \left( +73 \atop -84 \right) \cong 0.30 \pm 0.1,
\label{2}
\ee
and, with $\eta_0 = 0.967 (73) \cong 1.0$,
\be
\hat{b}_0 (Q^2) = \frac{1}{(Q^2)^{\eta_0}} \cong \frac{1}{Q^2}.
\label{3}
\ee
According to (\ref{1}) and (\ref{3}), in the tensor-Pomeron model,
for $Q^2 \gsim 20 GeV^2$, the photoabsorption cross section fulfills
the proportionality \cite{Ewerz}
\be
\sigma_{\gamma^*p} (W^2,Q^2) \propto \frac{(W^2)^{\epsilon_0}}{Q^2}
\label{4}
\ee
with $\epsilon_0 \cong 0.30$ from (\ref{2}).

In ref. \cite{DIFF2000}, in a phenomenological fit to the photoabsorption
experimental data, for $x \lsim 0.1$ and $0 \le Q^2 \lsim 1000 GeV^2$, technically
by assuming a piecewise linear functional dependence of the cross section
on the variable $\eta (W^2,Q^2)$ introduced \cite{DIFF2000} via
\be
\frac{1}{\eta (W^2Q^2)} = \frac{\Lambda^2_{sat} (W^2)}{Q^2+m^2_0},
\label{5}
\ee
where
\be
\Lambda^2_{sat} (W^2) \propto (W^2)^{C_2},
\label{6}
\ee
we found
\be
\sigma_{\gamma^*p} (W^2,Q^2) = \sigma_{\gamma^*p} (\eta (W^2,Q^2)).
\label{7}
\ee
According to our fit, the photoabsorption cross section, including its
$Q^2 = 0$ limit, only depends on the single low-x scaling variable
$\eta (W^2,Q^2)$, rather than $W^2$ and $Q^2$ separately. The fit gave
\cite{DIFF2000}
\be
C_2 = 0.28 \pm 0.06
\label{8}
\ee
for the exponent $C_2$ in (\ref{6}), as well as $m^2_0 = 0.125 \pm
0.027~ {\rm GeV}^2$ for the mass parameter $m_0$, with $m^2_0 < m^2_\rho$,
as expected, where $m_\rho$ denotes the $\rho$-meson mass.

By inspection of the graphical representation of the experimental data
for the photoabsorption cross section as a function of $\eta (W^2,Q^2)$,
one reads off an explicit functional dependence that is approximately
given by \cite{DIFF2000}
\bqa
\sigma_{\gamma^*p} (W^2,Q^2) = & \sigma_{\gamma^* p} (\eta(W^2,Q^2)) \propto
\frac{1}{\eta(W^2,Q^2)} = \frac{(W^2)^{C_2}}{Q^2},\nonumber \\
& (Q^2 + m^2_0 \gg \Lambda^2_{sat} (W^2)),
\label{9}
\eqa
where $\Lambda^2_{sat} (W^2) \lsim 7 {\rm GeV}^2$ in the energy range, where
experimental data are available.

Comparison of (\ref{9}) with (\ref{4}), and of (\ref{8}) with (\ref{2}),
reveals that for $Q^2 \gsim 20 {\rm GeV}^2$ the tensor-Pomeron fit (1) from ref. \cite{Ewerz}
confirms the result of the phenomenological fit (\ref{9}) of
$\sigma_{\gamma^*p} (W^2, Q^2) \propto 1/\eta (W^2,Q^2)$ from ref. \cite{DIFF2000}.

A comment on the two-Pomeron fit of ref. \cite{Donnachie} is appropriate. In distinction 
from the model-independent fit of the $Q^2$ dependence in ref. \cite{Ewerz},
the $Q^2$ dependence of the photoabsorption cross section in ref. \cite{Donnachie}
is described by parameter-dependent analytic expressions in $Q^2$. The fit gave
\cite{Donnachie} an intercept of $\alpha_0 \cong 1.4$ or
\be
\epsilon_0 \cong 0.40,
\label{10}
\ee
significantly different from (\ref{2}).

It is gratifying that the result of the tensor-Pomeron fit \cite{Ewerz} for $Q^2 \gsim 20
{\rm GeV}^2$, with $\epsilon_0 \cong 0.30$, supports and confirms the 
fit in ref. \cite{DIFF2000}. The tensor-Pomeron fit, under the restriction to
$Q^2 \gsim 20 {\rm GeV}^2$, has rediscovered the empirical scaling law \cite{DIFF2000}
of $\sigma_{\gamma^*p} (W^2,Q^2) = \sigma_{\gamma^*p} (\eta (W^2, Q^2))$.

Scaling in $\eta (W^2,Q^2)$, including the observed specific functional dependence (\ref{9}) of
$\sigma_{\gamma^*p} (W^2,Q^2) \propto 1/\eta (W^2,Q^2) = \Lambda^2_{sat} (W^2)/Q^2$,
valid for sufficiently large $Q^2$, is a consequence of the color-dipole picture
\cite{DIFF2000}. For clarity and completeness, we elaborate on this essential point in
detail.

The color dipole picture (CDP) of deep inelastic scattering (DIS) at low x is based
on the transverse position-space representation
\bqa
\sigma_{\gamma^*_{L,T}} (W^2, Q^2) & = \int dz \int d^2 \vec r_\bot
\vert \psi_{L,T} (\vec r_\bot, z (1 - z), Q^2) \vert^2 \nonumber \\
& \sigma_{(q\bar q)p} (\vec r^{~2}_\bot, z (1 - z), W^2) 
\label{11}
\eqa
combined with the interaction of a
$q \bar q$ color dipole with the color field in the nucleon \cite{Cvetic},
\bqa
\sigma_{(q \bar q)p} (\vec r^{~2}_\bot, z(1-z), W^2) & = \int d^2 \vec l_\bot \tilde{\sigma}
(\vec l^{~2}_\bot, z(1-z), W^2) \nonumber \\
& \left(1-e^{-i \vec l_\bot \vec r_\bot}\right).
\label{12}
\eqa
In (\ref{11}) and (\ref{12}), $\vec r_\bot$ denotes the transverse interquark
distance, $z(1-z)$ with $0 \le z \le 1$ describes the quark-antiquark $(q \bar q)$ configuration,
and $\vec l_\bot$ the transverse gluon momentum absorbed by the $q \bar q$ state.

The representation (\ref{11}), including the $(q \bar q)p$ interaction (\ref{12}),
follows \cite{Cvetic} from off-diagonal vector dominance \cite{Sakurai}
upon taking into account the internal
structure of the
vector states, $V \equiv (q \bar q)^{J=1}$, in the $\gamma^* \to V$ transition,
and transforming from momentum space to transverse position space. The minus sign
in front of the exponential in (\ref{12}), in the pre-QCD era originated \cite{Sakurai}, \cite{Cvetic}
from the required consistency \cite{Sakurai} of scaling behavior in the timelike region of
$e^+ e^-$-annihilation, $e^+e^- \to$ hadrons, and in the spacelike region of deep
inelastic scattering (DIS) of $e^- p \to e^- +$ hadrons.

In QCD, the $(q \bar q)^{J=1}$ states act as color-dipole states, thus recognizing
the interaction (\ref{12}) as a consequence of the color-dipole nature of the
$(q \bar q)^{J=1}$ states. In the limit of vanishing dipole size, $\vec r^{~2}_\bot \to 0$,
a $(q \bar q)$-color-dipole state acts as a color-neutral object of vanishing interaction
with the color field in the proton (``color transparency''), while for large dipole size,
$\vec r^{~2}_\bot \to \infty$, the cross section has to remain finite and becomes hadronlike
(``saturation''), thus leading to (\ref{12}). A specific ansatz for the $(q \bar q)p$ cross
section fulfilling the requirements of color transparency and saturation is mentioned in
(\ref{28}) below.

One observes that perturbative QCD (pQCD) has not been mentioned in arriving at the 
interaction cross section (\ref{12}). In pQCD, the form of the interaction (\ref{12}) is
understood as a consequence of the color-gauge-invariant interaction of the $(q \bar q)$
dipole with the color field in the nucleon via exchange of (at least) two gluons. The
form of the position-space representation (\ref{11}) with (\ref{12}) from off-diagonal GVD,
from the point of view of pQCD is accordingly recognized as a consequence of the gauge-invariant interaction of a $(q \bar q)^{J=1}$-color-dipole state with the gluon field in
the proton. The representation of the photoabsorption cross section in (\ref{11}) and
(\ref{12}) is nevertheless a non-perturbative\footnote{In ref. \cite{Nachtmann},
``Towards a Non-Perturbative Derivation of the CDP'', in a very detailed field-theoretic
analysis the form of the photoabsorption cross section (\ref{11}) was re-established, by
assumption excluding a dependence of the dipole cross section on $z(1-z)$ in (\ref{11}).
The treatment in ref. \cite{Nachtmann}, rederiving (\ref{11}), while not incorporating
color-transparency and saturation, misses an essential element of the CDP.} 
one, otherwise it would not include and contain the smooth transition to the $Q^2 = 0$
photoproduction cross section.

It is the structure of the $(q \bar q)p$-interaction amplitude contained
in (\ref{12}) that is responsible for the CDP prediction of the $1/Q^2$ dependence
of the photoabsorption cross section, $\sigma_{\gamma^*p} \propto 1/Q^2$ (given in
(\ref{17}) below), that, according to the fit result (\ref{1}) with (\ref{3}), is
confirmed by the experimental data.

Upon introducing the variables $\vec r^{~\prime} = \sqrt{z(1-z)} \vec r_\bot$ and $\vec l^{~\prime}_\bot
= \vec l_\bot / \sqrt{z(1-z)}$, and the cross sections for longitudinally and transversely
polarized $(q \bar q)^{J=1}_{L,T}$ ($J = 1$, vector) states, for the color-dipole-proton cross
section (\ref{12}), one indeed finds \cite{DIFF2000,Kuroda} the two limiting cases of ``color
transparency'' for $\vec r^{~\prime 2}_\bot \to 0$,
\bqa
\sigma_{(q \bar q)^{J=1}_{L,T} p} (r^{~\prime 2}_\bot, W^2) & = \frac{1}{4}
\pi \vec r^{~\prime 2}_\bot \int d \vec l^{~ \prime 2}_\bot \vec l^{~\prime 2}_\bot
\bar \sigma_{(q \bar q)^{J=1}_{L} p} (\vec l^{~\prime 2}_\bot, W^2)\nonumber \\
\times \left\{ \begin{array}{l}
1, \\
\rho_W
\end{array} \right. 
& \left( \vec r^{~\prime 2}_\bot \ll
\frac{1}{\vec l^{~\prime 2}_{\bot~Max} (W^2)} \right),
\label{13}
\eqa
where
\be
\rho_W \equiv \frac{\int d \vec l^{~\prime 2}_{\bot} \vec l^{~\prime 2}_{\bot}
\bar \sigma_{(q \bar q)^{J=1}_T p} (\vec l^{~\prime 2}_{\bot}, W^2)}{\int
d \vec l^{~\prime 2}_{\bot} \vec l^{~\prime 2}_{\bot}
\bar \sigma_{(q \bar q)^{J=1}_L p} (\vec l^{~\prime 2}_{\bot}, W^2)},
\label{14}
\ee
and ``saturation'' for $\vec r^{~\prime 2}_\bot \to \infty$,
\bqa
\sigma_{(q \bar q)^{J=1}_{L,T} p} & (r^{~\prime 2}_\bot, W^2)  =  \pi \int d \vec l^{~\prime 2}_\bot
\bar \sigma_{(q \bar q)^{J=1}_{L,T} p} (\vec l^{~\prime 2}, W^2) \nonumber \\
& \equiv  \sigma_{L,T}^{(\infty)}
(W^2),~~~~\left(\vec r^{~\prime 2}_\bot \gg \frac{1}{\vec l^{~\prime 2}_{\bot~Max}(W^2)} \right).
\label{15}
\eqa
In (\ref{13}) and (\ref{15}), $\vec l^{~\prime 2}_{\bot~Max} (W^2)$ denotes the upper limit
of the range of integration over $d \vec l^{~\prime 2}_\bot$ that yields the dominant contribution
to the integrals. The parameter $\rho_W$ in (\ref{14}), where $W$ indicates a potential dependence
of $\rho = \rho_W$ on the energy $W$, determines the ratio of (the first moment of) the cross
sections for the scattering of transversely (helicity $\pm 1$) and longitudinally (helicity $0$)
polarized $(q \bar q)^{J=1}_{L,T}$ states on the proton. The deviation from $\rho_W = 1$ quantifies
the deviation of the $(q \bar q)^{J=1}$-proton cross section from helicity independence of
$(q \bar q)^{J=1}_L p = (q \bar q)^{J=1}_T p$.

The cross section $\sigma^{(\infty)}_{L,T} (W^2)$ in (\ref{15}) is a purely hadronic quantity that
fulfills hadronic unitarity, and it is at most weakly, logarithmically, dependent on the energy $W$.

The photoproduction cross section, as a consequence of the explicit form of the ``photon wave function''
$\psi_{L,T} (\vec r_\bot, z(1-z),Q^2)$
in (\ref{11}), for sufficiently large $Q^2 \gg 0$, receives contributions from only a finite range of
$\vec r^{~\prime 2}_\bot$ that is restricted by $\vec r^{~\prime 2}_\bot \le 1/Q^2$. 
For given large energy $W$ and sufficiently
large $Q^2$, such that $1/Q^2 \ll 1 / \vec l^{~\prime 2}_{\bot~Max} (W^2)$, the photoabsorption cross
section (\ref{11}) is determined by the color-dipole cross section for $\vec r^{~\prime 2}_\bot \ll 1
/ \vec l^{~\prime 2}_{\bot~Max} (W^2)$, i.e. in the color-transparency limit. Introducing the first
moment of the distribution in $\vec l^{~\prime 2}_\bot$ for $(q \bar q)^{J=1}_L p$ scattering,
\be
\Lambda^2_{sat} (W^2) \equiv \frac{\int d \vec l^{~\prime 2}_\bot \vec l^{~\prime 2}_\bot
\bar \sigma_{(q \bar q)^{J=1}_L p} (\vec l^{~\prime 2}_\bot , W^2)}{\int d \vec l^{~\prime 2}_\bot 
\bar \sigma_{(q \bar q)^{J=1}_L p} (\vec l^{~\prime 2}_\bot , W^2)},
\label{16}
\ee
the photoabsorption cross section (\ref{11}), upon inserting (\ref{13}) and employing (\ref{14}) and
(\ref{15}), becomes \cite{DIFF2000, Kuroda}
\be
\sigma_{\gamma^*_{L,T} p} (W^2,Q^2) \propto \sigma_L^{(\infty)} (W^2) \frac{\Lambda^2_{sat} (W^2)}{Q^2}
\left\{ \begin{array}{l}
1, \\
2 \rho_W
\end{array} \right. .
\label{17}
\ee
With the approximation of $\sigma_L^{(\infty)} (W^2) \cong {\rm const}$, and $\rho_W = \rho = {\rm const}$
due to invariance under Lorentz boosts, and specifying $\Lambda^2_{sat} (W^2) \propto (W^2)^{C_2}$, we indeed
recognize the result of the phenomenological fit (\ref{9}) (that agrees with the tensor-Pomeron
fit (\ref{1})) as a consequence of the CDP (\ref{11}) and (\ref{12}).
The energy dependence of the fit (\ref{9}), as $\Lambda^2_{sat} (W^2) \propto
(W^2)^{C_2}$, according to (\ref{17}) with (\ref{16}) is recognized as the energy dependence
of the effective transverse momentum of the gluon absorbed by the color-dipole $q \bar q$ state
in the $\vec r^{~\prime 2}_\bot \to 0$ limit.

While the fitted large-$Q^2$ energy dependence (\ref{1}) of the photoabsorption cross section
in the tensor-Pomeron model, as $(W^2)^{\epsilon_0}$, is consistent with the energy dependence
(\ref{9}) of the CDP, as $(W^2)^{C_2}$, its interpretation \cite{Ewerz} is entirely different.
The fitted energy dependence for $Q^2 \gsim 20 {\rm GeV}^2$, in the tensor-Pomeron model
is interpreted as evidence for the exchange of a hypothetical ``hard-Pomeron trajectory'' with
intercept $\alpha_0 (0) = 1 + \epsilon_0$, independent empirical evidence for this
conjectured trajectory being lacking. Independently of this interpretation, the agreement
of the fit (\ref{1}) with the  fit (\ref{9}) now explicitly being recognized as a consequence of the
CDP, explicitly confirms our above conclusion that the fit (\ref{1}) provides
additional empirical support for the previously empirically established validity 
\cite{DIFF2000} of the
CDP of DIS at low $x \lsim 0.1$.

\section{Consistency of pQCD with the CDP and the numerical value of $\mathbf{C_2 \cong 0.30}$}
For $Q^2 \gsim 20 {\rm GeV}^2$, in both, the fit (\ref{4}) and the fit (\ref{9}), the
$Q^2$ dependence of the photoabsorption cross section is given by
\be
\sigma_{\gamma^*p} (W^2,Q^2) \propto \frac{1}{Q^2}.
\label{18}
\ee
The structure function of the proton, $F_2 (x,Q^2) \cong Q^2 \sigma_{\gamma^*p} (W^2,Q^2)$,
accordingly only depends on $W^2$,
\be
F_2 (x,Q^2) = F_2 \left( W^2 = \frac{Q^2}{x}\right).
\label{19}
\ee
Explicitly, this $W$ dependence of the experimental data is shown in Fig. 1 taken
from ref. \cite{Kuroda}. A eye-ball {\it two-parameter} fit to the experimental
data in fig. 1 gives \cite{Kuroda}
\be
F_2 (W^2) = f_2 \left( \frac{W^2}{1 {\rm GeV}^2} \right)^{C_2},
\label{20}
\ee
where $f_2 \cong 0.063$ and $C_2 = 0.29$.

\begin{figure}
\hspace*{-0.5cm}
\includegraphics[width=9cm]{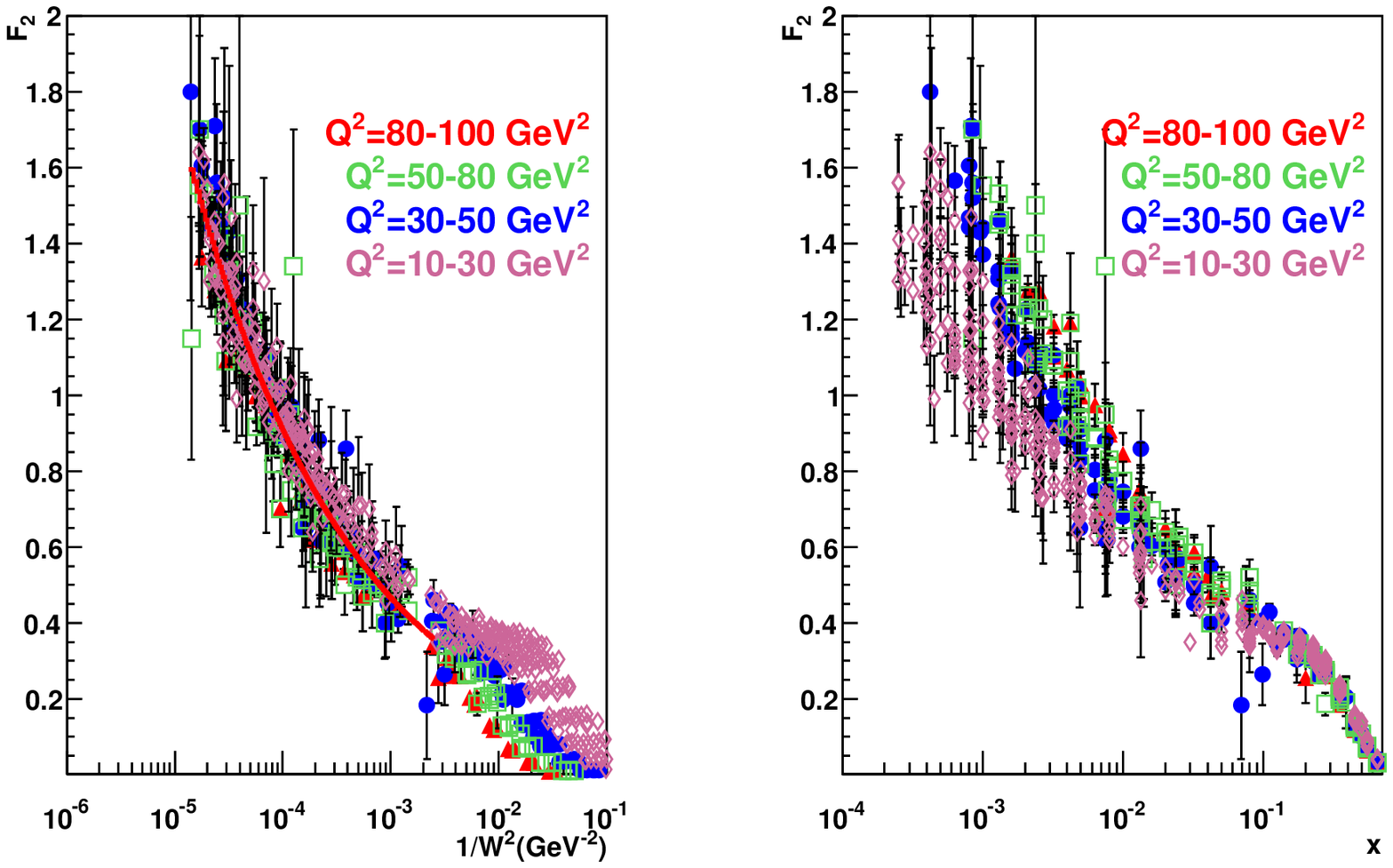}%
\vspace*{-0.5cm}
Fig.1a   \hspace{3.5cm} Fig.1b
\caption{In fig.1a we show the experimental data for $F_2(x\cong Q^2/W^2, Q^2)$ as a
function of $1/W^2$, and in fig.1b, for comparison, as a function of $x$. The
theoretical result 
based on (\ref{20}) is also shown in fig.1a.\label{Fig.1}}
\end{figure}

The observed $W^2$ dependence (\ref{20}) of the structure function
$F_2 = F_2 (W^2)$, interpreted as a consequence of the perturbative-QCD (pQCD) improved
parton model, implies a remarkable constraint \cite{Kuroda} on the magnitude of the exponent $C_2$.

In the pQCD improved parton model, the longitudinal structure function
$F_L (x,Q^2) = F_2 (x,Q^2) / (1+1/R)$, where $R$ refers to the longitudinal-to-transverse
photoabsorption-cross-section ratio, at a rescaled value $x \to \xi_L x$, for a wide range of different gluon distributions, is proportional to
the gluon distribution multiplied by $\alpha_s (Q^2)$, compare
ref. \cite{Martin}. Similarly, the logarithmic
derivative of $F_2 (x,Q^2)$, at a rescaled value of $x \to \xi_L x$, is proportional
to the gluon distribution \cite{Prytz}. Eliminating the gluon distribution, and adopting
the empirically supported dependence (\ref{20}) of $F_2 = F_2 (W^2)$, one finds
an evolution equation for $F_2 (W^2)$ that is given by \cite{Kuroda}
\be
(2 \rho_W + 1) \frac{\partial}{\partial ln W^2} F_2 \left( \frac{\xi_L}{\xi_2} W^2 \right)
= F_2 (W^2).
\label{21}
\ee
In (\ref{21}), we have replaced the longitudinal-to-transverse ratio of the photoabsorption
cross sections via $R = 1/2 \rho_W$, compare (\ref{17}).

Inserting $F_2 \propto (W^2)^{C_2}$, see (\ref{20}), from (\ref{21}),
we obtain the constraint \cite{Kuroda}
\be
(2 \rho_W + 1) C_2 \left( \frac{\xi_L}{\xi_2} \right)^{C_2} = 1.
\label{22}
\ee
The constraint (\ref{22}) connects the exponent $C_2$ of the
power-law dependence (\ref{20}) of $F_2 = F_2 (W^2)$ with the value
of $\rho_W$ that quantifies the deviation from helicity independence,
$\rho_W = 1$, of $(q \bar q)^{J=1}$ scattering from the proton
according to (\ref{14}). Constancy of $C_2$, according to (\ref{22}),
implies constancy of $\rho_W = \rho$. From the point of view of pQCD,
the constancy\footnote{A refined analysis \cite{Kuroda} of the CDP leads to a correction
of $R=1/2 \rho$ that implies a decrease of $R \propto 1/\eta (W^2,Q^2) =
\Lambda^2_{sat} (W^2)/Q^2$ in the limit of very large values of $Q^2$
that reach the limit of $x \cong Q^2/W^2 \to 0.1$.}
of $\rho$ implies \cite{Kuroda} a gluon
distribution multiplied by $\alpha_s (Q^2)$ that coincides with the
$q \bar q$ sea distribution.

Expanding the exponential in (\ref{22}) (to first order in $C_2$), and
solving for $C_2$ yields
\be
C_2 \cong \frac{1}{2 \rho + 1} \frac{1}{\left(1-\frac{1}{2 \rho +1} ln
\frac{\xi_2}{\xi_L} \right)}.
\label{23}
\ee
The ratio of $\xi_2 / \xi_L$ in (\ref{22}) and (\ref{23}) is given
by \cite{Martin, Prytz} $\xi_2 / \xi_L = 1.25$.

\begin{table}[h]
\caption
{The value of $\rho$ determines the ratio
of the cross sections for transversely, $(q \bar q)^{J=1}_T$, and
longitudinally polarized, $(q \bar q)^{J=1}_L$, states on the proton,
see (\ref{14}). A value of $\rho \not= 1$ corresponds to a deviation
of $R$ from the value of $R = 0.5$ corresponding to helicity independence
of $(q \bar q)^{J=1} p$ scattering. Theoretical values of $C_2$ from
(\ref{23}) are compared with experimental ones. For $C_2 \vert_{Exp} = 0.27 \pm 0.1$,
see (\ref{30}) below.}
\begin{ruledtabular}
\begin{tabular}{|c|c|c|c|l|}
$\rho$ & $R = \frac{1}{2 \rho}$ & $\frac{1}{2 \rho + 1}$ & $C_2$ & $C_2 \vert_{Exp}$ \\
\hline
1 & 0.5	 &$\frac{1}{3} \cong 0.33$ & 0.360 & \\
\hline
 &  & & & $0.30 \pm 0.01$ (ref.1)\\
\rb{$\frac{4}{3}$} & \rb{0.375} & \rb{$\frac{3}{11} \cong 0.27$} & \rb{0.290} & $0.27 \pm 0.01$ (ref.3)\\
\hline
2 & 0.25 & $\frac{1}{5} = 0.20$ & 0.209 & \\
 \end{tabular}
 \end{ruledtabular}
 \end{table}

In Table 1\footnote{The ``rigorous upper limit'' of $R \le 0.37248$ from the
``standard color-dipole model of low x DIS''\cite{Ewerz, Nachtmann}, excluding
$\rho = 1 (R = 1/2)$ and even $\rho = 4/3 = 0.375$ in Table 1, depends \cite{Schi}
on the ad hoc assumption contained in the ``standard color-dipole model'' \cite{Ewerz,
Nachtmann} of excluding \cite{Nachtmann} a $z(1-z)$ dependence of the $(q \bar q)p$
cross section $\sigma_{(q \bar q)p}$ in (\ref{11}),
compare footnote 2. Fits to experimental data test this underlying assumption.
A violation of the ``rigorous upper limit'' by experimental data can be taken
care of by allowing for a dependence  on $z(1-z)$, see e.g. the ansatz (\ref{28})
that implies $\rho = 1$ (i.e. helicity independence of $(q \bar q)^{J=1} p$ scattering,
compare (\ref{14}))
and $R = 1/2\rho = 0.5$ in violation of the bound.},
we show the results for $C_2$ from (\ref{23}) for different
values of $\rho$ and $R=1/2 \rho$. We comment on the results in Table 1 as follows:
\begin{itemize}
\item[i)] The different dependence on the configuration, $z (1-z)$,
of longitudinally and transversely polarized $(q \bar q)^{J=1}$ states,
via the uncertainty principle, implies an enhanced transverse size of
transversely polarized relative to longitudinally polarized $(q \bar q)^{J=1}$
states. This estimate within the CDP implies $\rho = 4/3$ \cite{Kuroda}.
Consistency of the CDP with pQCD, from (\ref{23}), predicts $C_2 = 0.29$,
in agreement with the observed value of $C_2\vert_{Exp} = \epsilon_0 \cong 0.30$
from (\ref{2}) and $C_2 \vert_{Exp} = 0.28 \pm 0.06$ from (\ref{8}) and
$C_2 \vert_{Exp} = 0.27 \pm 0.01$ from (\ref{30}) below. We conclude that the consistency between
the CDP and pQCD is empirically established.

\item[ii)] A value of $\rho = 1$, assuming helicity independence
of $(q \bar q)^{J=1} p$ scattering, in the approximation of ignoring the transverse-size
enhancement discussed in i), according to the consistency condition (\ref{23}) implies 
$C_2 = 0.360$, which is excluded
by the measured value of $C_2\vert_{Exp} = \epsilon_0 \cong 0.30$.
Concerning the direct experimental determination of $R = 1/2 \rho$, see
below.

\item[iii)] A value of a transverse-to-longitudinal enhancement by a factor of
$\rho = 2$, or $R = 1/2 \rho = 0.25$ requires $C_2 = \epsilon_0 = 0.21$, and is
definitely excluded by $C_2 \vert_{Exp} \cong 0.30$.

\item[iv)] We add a comment on the direct measurement of the ratio $R = 1/2 \rho$
for $Q^2 \gg 0$. First results from the H1 and ZEUS collaborations \cite{H1, ZEUS} showed
agreement \cite{Kuroda}  with the prediction of
\bqa
F_L = \frac{1}{1 + \frac{1}{R}} & F_2 = \frac{1}{1+2 \rho} F_2 = 0.27 F_2,\nonumber \\
& (\rho = 4/3).
\label{24}
\eqa
More recent measurements showed a smaller value of R \cite{Andreev}. The more
indirect determination of $R$ from the two-Pomeron fit to the reduced cross sections
led to values roughly between $R = 0.35$ and $R = 0.5$ \cite{Ewerz}. It is very
unfortunate that precision measurements on this very important ratio
$R$ will not become available in the near-by
future. 
\end{itemize}

\section{Low values of $\mathbf{Q^2}$, the photoproduction limit of $\mathbf{Q^2 \to 0}$}
We come to the fourth part of this article, the examination of the low-$Q^2$ region
of $Q^2 \lsim 20 {\rm GeV}^2$, including the photoproduction limit of $Q^2 = 0$.

In the tensor-Pomeron model, with decreasing $Q^2$, the contribution of the soft
Pomeron (plus $f_{2R}$ Reggeon) to the total photoabsorption cross section becomes
increasingly important. In the $Q^2 = 0$ photoproduction limit, the hard Pomeron
yields \cite{Ewerz} a vanishing contribution. The soft-Pomeron contribution behaves
as \cite{Ewerz}
\be
\sigma_{\gamma^* p} (W^2,Q^2) \propto (W^2)^{\epsilon_1},~~{\rm for}~(Q^2 \to 0),
\label{25}
\ee
where
\be
\epsilon_1 = 0.0935 \left( {+76 \atop -64} \right) \cong 0.094,
\label{26}
\ee
close to $\epsilon_1 = 0.096$ from ref. \cite{Donnachie}.

In the CDP, with decreasing $Q^2 \to 0$ or increasing $1/\eta (W^2,Q^2)$, 
the photoabsorption cross
section receives contributions from an increasingly larger dipole
size $\vec r_\bot^{~\prime 2}$. The photoabsorption cross section
(\ref{11}) is determined by the $\vec r^{~\prime 2}_\bot \to \infty$
saturation limit (\ref{15}) of the dipole cross section (\ref{12}). 
A detailed examination, based on (\ref{15}), yields \cite{Kuroda}
\bqa
\sigma_{\gamma^*p} (W^2, Q^2) & \propto \sigma_T^{(\infty)} (W^2) ln
\frac{\Lambda^2_{sat} (W^2)}{Q^2 + m^2_0}, \nonumber \\
&\left( Q^2 + m^2_0 \ll \Lambda^2_{sat}
(W^2) \right).
\label{27}
\eqa
The transition from $Q^2 \gg \Lambda^2_{sat} (W^2)$ in (\ref{17}) to
$Q^2 \ll \Lambda^2_{sat} (W^2)$ in (\ref{27}) occurs by transition
from $\sigma_{\gamma^* p} (W^2,Q^2) \propto 1/\eta (W^2,Q^2)$ in
(\ref{17}) to $\sigma_{\gamma^*p} (W^2,Q^2) \propto ln \left( 1 / \eta 
(W^2,Q^2) \right)$ in (\ref{27}). In distinction from the two-Pomeron
model, no independent additional parameter must be introduced in
this smooth transition from $Q^2 \gg 20 {\rm GeV}^2$ to $Q^2 = 0$.
The cross section (\ref{27}) fulfills \cite{Kuroda} the Froissart bound
$\sigma_{\gamma^*p} \propto ln^2 W^2$.\footnote{The photon makes a transition
to (quasi-asymptotic) on-shell $(q \bar q)^{J=1}$ states, see e.g.
ref. \cite{KuSchi} for a concise representation of this point. For e.g. $Q^2 \to 0$, the range of masses of contributing
$(q \bar q)^{J=1}$ states is essentially given by transitions $\gamma
\to (q \bar q)^{J=1}$ to low-lying $(q \bar q)^{J=1}$ on-shell color-dipole
states representing the low-lying vector mesons, $\gamma \to (\rho^0, w, \phi)$.
(For an early representation of the $(\rho^0, w, \phi)$ contributions by a 
massive $J=1$ continuum compare ref. 5.) Accordingly, $\sigma_{\gamma p}
(W^2) \propto \sigma_{(\rho^0,w,\phi)p} (W^2) \propto ln^2 W^2$, where the
second step rests on the $ln^2 W^2$ dependence of the hadronic cross section
$\sigma_{(\rho^0,w,\phi)} (W^2)$ in the high-energy limit. \cite{Heisenberg}.
In (\ref{27}) and in (\ref{28}), we have $\sigma^{(\infty)} (W^2) \propto
ln W^2$. Compare also ref. \cite{KuSchi} for a discussion of the $W$ dependence
of $\sigma^{(\infty)} (W^2)$ and its experimental determination.
The fact that $\gamma^*$ is not an asymptotic state \cite{Ewerz} is
recognized as becoming irrelevant with respect to the high-energy limit of the
photo-absorption cross section as
soon as the underlying (empirically supported \cite{DIFF2000, Sakurai}) dynamical
mechanism of $\gamma^* \to (q \bar q)^{J=1}$ with subsequent on-shell hadronic
$(q \bar q)^{J=1} p$ scattering is taken into account.
For successful fits to the measured total cross sections for
$\gamma^*p$, as well as $\gamma p, \pi^\pm p, pp, \bar p p$, scattering, based
on Froissart-bound saturation $(ln^2 W^2)$ rather than Regge theory $\left( (W^2)^{\alpha_j (0)-1}\right)$,
compare ref. \cite{Block}}.

The considerations on the CDP so far were exclusively obtained by analysing
the general expressions (\ref{11}) and (\ref{12}). A detailed interpolation
between the limits of $1/\eta (W^2, Q^2)$ for $\eta (W^2,Q^2) \gg 1$, and
$ln \left(1 / \eta (W^2,Q^2) \right)$ for $\eta (W^2,Q^2) \lsim 1$, requires
a specific ansatz for the dipole cross section in (\ref{11}).
The fit in ref. \cite{DIFF2000}
was based on the ansatz
\bqa
& \sigma_{(q \bar q)p} (\vec r_{\bot}, z (1-z), W^2) \nonumber \\
& = \sigma^{(\infty)} (W^2) \left( 1 - J_0 \left( r_\bot z(1-z) \Lambda^2_{sat}
(W^2) \right) \right),
\label{28}
\eqa
where $\sigma^{(\infty)} (W^2)$ is of hadronic size and logarithmically dependent
on $W^2$. The ansatz (\ref{28}) contains color transparency (\ref{13}) and
saturation (\ref{15}). The dependence on $r^\prime_\bot = r_\bot z (1-z)$
implies \cite{Kuroda} $\rho = 1$ and $R = 1/2 \rho = 1/2$ for $Q^2 \gg 0$\footnote{A refined ansatz \cite{Kuroda} for the dipole interaction incorporates $\rho = 4/3$}.

The fit to the total photoabsorption cross section (\ref{11}) upon insertion of
(\ref{28}) gave \cite{DIFF2000}
\be
\Lambda^2_{sat} (W^2) = C_1 \left( \frac{W^2+W^2_0}{1 {\rm GeV}^2} \right)^{C_2}
\simeq C_1 \left( \frac{W^2}{1 {\rm GeV}^2} \right)^{C_2},
\label{29}
\ee
where
\bqa
C_2 & = & 0.27 \pm 0.01, \nonumber \\
C_1 & = & 0.34 \pm 0.05 {\rm GeV}^2,
\label{30}
\eqa
as well as
\bqa
m^2_0 & = 0.16 \pm 0.01 {\rm GeV}^2, \nonumber \\
W^2_0 & = 882 \pm 246 {\rm GeV}^2.
\label{31}
\eqa
One finds $2 {\rm GeV}^2 \lsim \Lambda^2_{sat} (W^2) \lsim 7 {\rm GeV}^2$
for the HERA energy range of $30 {\rm GeV} \lsim W \lsim 300 {\rm GeV}$. The
result (\ref{30}) is consistent with, and improves the accuracy of the
phenomenological fit (\ref{8}), and it is consistent with the tensor-Pomeron fit
(\ref{2}). For the fit, the hadronic cross section $\sigma^{(\infty)} (W^2)$
was consistently expressed \cite{DIFF2000, Kuroda} in terms of a fit to the $Q^2 = 0$ photoproduction
cross section. In addition to this input of a fit to the $Q^2 = 0$ photoproduction
cross section, essentially only three independent
fit parameters $C_2, C_1$ and $m^2_0$, have been used for a successful 
representation of the body of the experimental data on low-x DIS, including
the transition to $Q^2 = 0$ photoproduction. Compare the previous Section for a
theoretical prediction of the value of $C_2$.

\section{Conclusions}
We end with the following conclusions:
\begin{itemize}
\item[i)] The hard-Pomeron-fit result \cite{Ewerz} of $\sigma_{\gamma^*p}
(W^2,Q^2) \propto (W^2)^{\epsilon_0}/(Q^2)^{\eta_0}$ for $Q^2 \gsim 20 {\rm GeV}^2$,
where $\epsilon_0 = 0.3008 \left( {+73 \atop{-84}} \right) \cong 0.30 \pm 0.1$
and $\eta_0 = 0.967 (73) \cong 1.00 \pm 0.04$, based on the assumption of
a power-like $W^2$ dependence and 17 parameters for a model-independent
parametrization of the $Q^2$ dependence, confirms the prediction \cite{DIFF2000}
of the CDP of $\sigma_{\gamma^*p} (W^2,Q^2) \propto 1/\eta (W^2,Q^2) =
\Lambda^2_{sat} (W^2) /Q^2$ with the {\it predicted} $1/Q^2$ dependence and the fitted $W^2$ dependence of $\Lambda^2_{sat}
(W^2) \propto (W^2)^{C_2}$, where $C_2 = 0.27 \pm 0.01$.
\item[ii)] Requiring consistency of the CDP and pQCD for sufficiently large
$Q^2 \gsim 10~{\rm GeV}^2$ yields the {\it prediction} of $C_2 = 0.29$ 
for the exponent $C_2$ that is
confirmed by the CDP fit result of $C_2 = 0.27 \pm 0.01$ and the hard-Pomeron-fit
result of $\epsilon_0 \cong 0.30 \pm 0.1$.
\item[iii)] In distinction from the tensor-Pomeron approach, in the CDP, the transition from
$Q^2 \gsim 20 {\rm GeV}^2$ to $Q^2 \lsim 20 {\rm GeV}^2$, including the
$Q^2 = 0$ photoproduction limit, is obtained by a smooth transition
from $\sigma_{\gamma^*p} \propto 1/\eta (W^2,Q^2)$ to $\sigma_{\gamma^*p}
\propto ln (1/\eta (W^2,Q^2))$, without the necessity for introducing an
additional fit parameter.
\item[iv)] The essential new element of DIS compared to photoproduction consists of
the additional degree of freedom of the virtuality, $Q^2$, of the (virtual) photon.
The $Q^2$ dependence of the photoabsorption cross section in the CDP is 
{\it uniquely predicted} as a
consequence of the mass-dispersion relation of Generalized Vector Dominance (GVD)
formulated some 50 years ago. In the tensor-Pomeron-Regge approach this $Q^2$ {\it dependence
is fitted} by a large number of parameters to allow for a representation of the energy
dependence of the photoabsorption experimental data in terms of a linear superposition
of a power-law ansatz associated with, respectively, a soft and a hard Pomeron trajectory.
Additional independent empirical information on the (hard) Pomeron trajectory is lacking
so far, however.
\end{itemize}

\begin{acknowledgments}
The author thanks Kuroda-san for useful remarks on the preliminary version
of this paper.
\end{acknowledgments}

\end{document}